\documentclass[review]{elsarticle}

\makeatletter
\def\ps@pprintTitle{%
 \let\@oddhead\@empty
 \let\@evenhead\@empty
 \def\@oddfoot{\centerline{\thepage}}%
 \let\@evenfoot\@oddfoot}
\makeatother

\usepackage{lineno,hyperref}

\usepackage[table,xcdraw]{xcolor}
\usepackage{pifont}
\usepackage{graphicx}

\modulolinenumbers[5]

\journal{}

\usepackage{float}
\usepackage{xcolor}
\usepackage[english]{babel}
\begin{document}

\begin{frontmatter}

\title{A Cost-Effective, Scalable, and Portable IoT Data Infrastructure for Indoor Environment Sensing}

\author[anik_address]{Sheik Murad Hassan Anik}

\author[gao_address]{Xinghua Gao\corref{emails}}

\author[anik_address]{Na Meng}

\author[gao_address]{Philip R Agee}

\author[gao_address]{Andrew P McCoy}

\cortext[emails]{xinghua@vt.edu}

\address[anik_address]{Department of Computer Science, Virginia Polytechnic Institute and State University, Blacksburg, VA 24061, USA}
\address[gao_address]{Myers-Lawson School of Construction, Virginia Polytechnic Institute and State University, Blacksburg, VA 24061, USA}







\begin{abstract}
The vast number of facility management systems, home automation systems, and the ever-increasing number of Internet of Things (IoT) devices are in constant need of environmental monitoring. Indoor environment data can be utilized to improve indoor facilities and better occupants’ working and living experience, however, such data are scarce because many existing facility monitoring technologies are expensive and proprietary for certain building systems, such as building automation systems, energy management systems, and maintenance systems. In this work, the authors designed and prototyped a cost-effective, distributed, scalable, and portable indoor environmental data collection system, Building Data Lite (BDL). BDL is based on Raspberry Pi computers and multiple changeable arrays of sensors, such as sensors of temperature, humidity, light, motion, sound, vibration, and multiple types of gases. The system includes a distributed sensing network and a centralized server. The server provides a web-based graphical user interface that enables users to access the collected data over the Internet. To evaluate the BDL system's functionality, cost-effectiveness, scalability, and portability, the research team conducted a case study in an affordable housing community where the system prototype is deployed to 12 households. The case study results indicate that the system is functioning as designed, costs about \$3500 to sense 48 building zones (about \$73 per zone) and provides 12 types of indoor environment data, is easy to scale up, and is fully portable. 
\end{abstract}

\begin{keyword}
Internet of Things, Sensing Network, Facilities Management, Indoor Environment Sensing, Raspberry Pi
\end{keyword}

\end{frontmatter}


\section{Introduction} \label{introduction}

Buildings are one of the most important aspects of human life, and people spend 87\% of their time inside buildings \cite{mujan2019influence}. Buildings have both direct and indirect impacts on human health, comfort, and productivity \cite{mujan2019influence, berglund1992effects,tang2020interactions, tham2016indoor}. Maintaining optimal conditions for a better experience is critical for built environments. New techniques and methods are becoming available for this purpose with the developments of the Internet of Things (IoT) technologies \cite{hui2017major, minoli2017iot}. For example, an air quality monitoring system can benefit occupants in many ways from detecting harmful gases to monitoring oxygen levels. There are systems, such as the ones presented in Zakaria \textit{et al}. \cite{zakaria2018wireless} and Marques \textit{et al}. \cite{marques2019cost}, that provide indoor air quality monitoring. However, these systems are dedicated for air quality monitoring only without the capability of extension to other features such as temperature or humidity monitoring. Another important application of indoor environment sensing system is to support the optimization of building energy consumption. Buildings consume about 40\% of total consumed energy, and half of that goes to Heating, ventilation, and air conditioning (HVAC) systems \cite{ekwevugbe2013real}. A significant portion of this consumed energy is wasted \cite{meyers2010scoping}, and using a proper monitoring and control system can reduce this waste significantly   \cite{ekwevugbe2013real}.


Even though many building systems already incorporate proprietary networks of sophisticated sensors and devices, they normally have very limited inter‐system connectivity or exposure to the larger networks of IoT devices, which hinders the establishment of a comprehensive IoT data infrastructure for building automation and management. 

There are four major challenges in establishing a comprehensive IoT data infrastructure for buildings. First, the high cost of the sensing devices is a major reason why building owners and managers are normally reluctant to install a data system in the first place. For example, the average cost per square foot to outfit a facility with a building automation System (BAS) is \$2.30, in 2017, which means it will cost about \$230,000 on BAS alone for a typical 100,000 square foot building \cite{readyone2017}. Moreover, building systems such as BAS will not last for the entire life cycle of a building, and there will be upgrade fees for about every ten years or less. Second, the installation of building systems can be a challenge for existing buildings. The best time to install a building system, with numerous wires and devices, is during construction. After a building starts to operate, installing additional systems or changing the existing ones are usually disruptive to occupants. 

The third challenge is that most of the data generated by commercial building systems are self-proprietary because most of the existing building systems are lacking interoperability. Typically, there is a lack of means to use the data collected in one system for another system's functions. Different systems need to generate the indoor environment data they need, and the data are not in a format that can be used for other purposes, which significantly limits the potential of smart building innovations. Scalability is the fourth challenge. For many building systems, after the initial installation during construction, it is usually difficult to modify the system by adding or removing features, or increasing the coverage to other building spaces. The system's functions and scale, including what kind of indoor environment data to generate, are usually settled by the time of initial installation. Scaling up the system may cost more than installing a brand new system that can cover the required functions and spaces. In addition, current building systems are lacking portability. Usually, after a system is installed in a building, it normally cannot be ported to a different location later on. It is very challenging to take out the system from the installed location and place it in another location without extensive engineering and re-construction. 

In this paper, we present a cost-effective, scalable, portable, and distributed indoor environment sensing system, Building Data Lite (BDL), to address the building data availability challenges, and thus, to provide a means for establishing the IoT data infrastructure of a smart building. To collect indoor environment data, the system has three types of components: sensing nodes, central nodes, and inter-node connections. The sensing nodes are an array of digital and analogue sensors to collect data from the surroundings. Each sensing node has its local database. A central node is a web-based central server that integrates data from all local databases and visualizes data on a website. The inter-node connections link sensing nodes with the central node. The system is scalable such that each sensing node's function (what data to generate) is customizable, and it is easy to add sensing nodes to or remove nodes from an established system. The system also includes a web-based Graphical User Interface (GUI) for users to extract, view, and analyze the generated data. Industry $4.0$ \cite{AHELEROFF2020101043} introduced a system that enables appliances such as refrigerators or air-conditioners to capture relevant data and pass it to a cloud storage. The BDL system, on the contrary, uses dedicated sensing nodes to capture targeted data through connected sensors. This research proposes an innovative means of establishing cost-effective, scalable, and portable IoT infrastructure for indoor environment data generation, integration, processing, and presentation. The research team developed a prototype of the BDL system to demonstrate how it overcomes the limitations of existing building systems regarding indoor environment sensing in the following manners:

a) Cost: The BDL system is built with mini computer Raspberry Pi and compatible sensor modules. The cost of each sensing node is limited to \$40 to \$60, depending on the sensing requirements. Typically, each sensing node can cover one building zone, and therefore, for a building with 100 zones, the entire system only costs about \$5,000. The BDL system is fairly affordable for indoor environment sensing requirements by smart building research and innovations. 

b) Installation: The distributed sensing nodes (Raspberry Pi computers) of the BDL system are connected with a centralized server via wireless communication. Each node can capture multiple indoor environment values of a particular building space. The installation of the sensing nodes requires only power supply and Internet/WiFi connection.

c) Comprehensiveness: The BDL system uses multiple sensors to collect different environment data simultaneously, and establish a comprehensive database that can be used for developing multiple smart building applications. The developed prototype can capture data related to temperature, humidity, light, sound, motion, vibration, flame, and three types of gas.

d) Scalability: The BDL system is customizable, and the users have the option to add or remove devices (Raspberry Pis), sensor modules, and functions.

e) Portability: The nodes of BDL system are connected to the central server via local WiFi or the Internet, which makes them portable to any location with power supply and under the coverage of wireless network. The sensing nodes can be removed from one location and then deployed in another location without any engineering on the building. 

f) Open-source: The source code of BDL has been published by the authors via GitHub \cite{AnikGitHub2021}. The BDL also provides an open-source, integrated database system that makes it easier for other systems to utilize the generated data. 


This paper is structured as follows: Section 2 reviews relevant studies conducted in the indoor environment sensing field. Section 3 presents the BDL system's overall design. Section 4 demonstrates the prototype development. Section 5 presents the case study in which the BDL system is deployed in 12 households. Section 6 discusses the challenges encountered and corresponding solutions, potential use cases of the BDL system, and the current limitations and future research directions. Section 7 concludes the research.

\section{Literature Review} \label{related_works}
The first step to make a building “smart” is to establish the data infrastructure. Data collection has always been a challenge in the smart building research domain \cite{linder2017big}. Wireless sensor network technologies have been implemented as solutions to this challenge. However, there is still a lack of cost-effective means to generate, collect, and process the ubiquitous indoor environment data in existing buildings. In this section, related works in the domain of indoor environmental sensing system are discussed. The innovation of the proposed BDL system lies in comprehensively achieving cost-effectiveness, portability, scalability, and generalizability. Therefore, the related works are discussed regarding these properties.     

Ferdoush \textit{et al}. \cite{ferdoush2014wireless} presented a wireless sensor network system using open-source hardware platforms, Arduino, Raspberry Pi, and XBee module. The system is low-cost and scalable both in terms of the type of sensors and the number of sensor nodes. It is suited for a wide variety of applications related to environmental monitoring. Each sensing node centers around an Arduino board, which incorporates multiple sensors. Ferdoush \textit{et al}. \cite{ferdoush2014wireless} demonstrated their prototype by collecting only temperature and humidity data. The Arduino board incorporates a XBee module to communicate with the base station. The base station is based on a Raspberry Pi which is connected to the Internet via a router. The XBee module on the base station works as a coordinator device which has a limitation of supporting maximum 10 sensing nodes. The system is portable in the sense of placing the sensing nodes with the approximation of the base station. 


Gross \textit{et al}. \cite{gross1984monitoring} conducted a study on the usage and condition of the emergency room at Soroka Hospital Center. The experiment had a fixed system setup in terms of the portability and scalability. The study has found that the adult respiratory conditions are likely to occur on days with high total particulates (TSP) and ``respirable'' particulates (RSP). A significant difference was found in RSP for adults (not in TSP), when the mean values were compared for asthma and shortness of breath days against normal days. Natural dust is the more likely cause of this association. The authors conclude that emergency room monitoring should be useful in locations with high levels of man-made pollutants. Truong \textit{et al}. \cite{truong2017iot} proposed an IoT system that can capture environmental data of rural crop fields, and store the data in the cloud. The authors applied machine learning on the collected data to analyze environmental conditions for harmful fungal diseases in local crop fields. The study focused on providing environmental data in rural crop fields for the detection and management of fungal diseases. This study is more of an application of a IoT based sensing network.



To lower energy consumption, Alva \textit{et al}. \cite{alva2012design} used video-based human occupancy sensing to optimize the lighting strategy. This study involves the development of a human occupancy sensing system in MATLAB and the hardware for lighting controls. Video cameras in this system can be placed in a portable fashion but the system remains purpose-specific, and the cameras significantly increase the system's cost.  


Kumar \textit{et al}. \cite{kumar2014android} proposed a low-cost Smart Living System, which uses an Android-based user interface for control of home appliances. This model integrates temperature and humidity sensor, DHT11, sound sensor, LM393, and gas sensor, MQ135, with Raspberry Pi 3B microcontroller board. The authors used an external GPRS module to connect their system with the internet. The paper presents an experiment on monitoring the air quality, sound, temperature, and humidity detection. The system uploads the collected data to a cloud server, which decides whether the pollution level has crossed a certain threshold. 


Zakaria \textit{et al}. \cite{zakaria2018wireless} developed an air quality sensing system by utilizing MQ-135 gas sensor, coupled with the temperature and humidity sensor DHT-22. The system can detect gaseous components such as NH3, NOx, alcohol, and benzene, along with sensing temperature and humidity. Coleman \textit{et al}. \cite{coleman2017examining} developed a low-cost indoor air quality sensing system capable of measuring volatile organic compounds (VOCs) and other gaseous concentrations while monitoring temperature and humidity. They identified the Indoor Air Quality (IAQ) matrix as composed of CO2, VOC, CO, PM, HCHO and NO2. Sahal \textit{et al}. \cite{saha2018raspberry} proposed a Raspberry Pi v3 based model to capture and monitor gas and sound properties from the environment. It uses a LM393 sound sensor, MQ135 gas sensor, DHT11 temperature and humidity sensor, and a GPRS module. The objective of this research is monitoring the air quality and measuring noise intensity to mitigate sound pollution. Marques \textit{et al}. \cite{marques2019cost} provided an IoT based real-time indoor air quality monitoring system named iAir. It features an ESP8266 with ATMega168PA MCU as communication and processing unit. The system utilizes MICS6814 sensor to detect gases such as Carbon Mono-oxide, Nitrogen Dioxide, Ethanol, Methane, and, Propane. It uses WiFi connectivity and smartphone application to provide data access and real-time notification. 

The website \cite{Cayenne2020} provides an online platform to make IoT projects featuring both Arduino and Raspberry Pi with just drag and drop. A list of sensors is provided in the system for users to use. Custom codes can also be injected if desired. It also provides the user a dashboard to access and monitor the data collected by the sensors connected in their system. The system helps individuals to create custom projects with user specific requirements.

The systems proposed in \cite{zakaria2018wireless, marques2019cost, coleman2017examining, saha2018raspberry} are focused on specific task of air quality monitoring with a set of fixed sensors, lacking the feature of scalability. The system in \cite{alva2012design} is also task-specific but significantly more expensive than the other works mentioned here. The systems proposed in \cite{ferdoush2014wireless, kumar2014android, Cayenne2020} offer the property of cost-effectiveness, portability, and scalability to some extent. The BDL system proposed in our work aims in attaining cost effectiveness, code-free scalability, portability, while keeping the usability of the system more generic and task independent. BDL's architecture is similar to the one presented by Aheleroff \textit{et al}. \cite{aheleroff2021digital}, which describes the system of mirroring digital representatives of physical assets with two-way dynamic mapping. In respect to \cite{Cayenne2020}, BDL provides a complete code-free sensing network with both local and central database whereas \cite{Cayenne2020} can be used to create functional sensing nodes. The general concept of code-free sensing network system can help research work across researchers of different backgrounds because data regarding building environments can be useful in multiple academic areas, in which not all researchers have a background of programming. 

\section{The System Design of BDL} \label{research_method}


Unlike previous studies that develop a sensing system for a particular use case, this research is focusing on how to develop a more generic-purpose, cost-efficient, scalable, portable, and distributed indoor environment data sensing system that can be used to establish the data infrastructure for smart building innovations. This section demonstrates the system design of BDL by describing the system architecture, central server, sensing node, database configuration and data transmission, network configuration, data integration, and graphic user interface.



\subsection{System Architecture}

\begin{figure}[t]
\centering
\includegraphics[width=\textwidth]{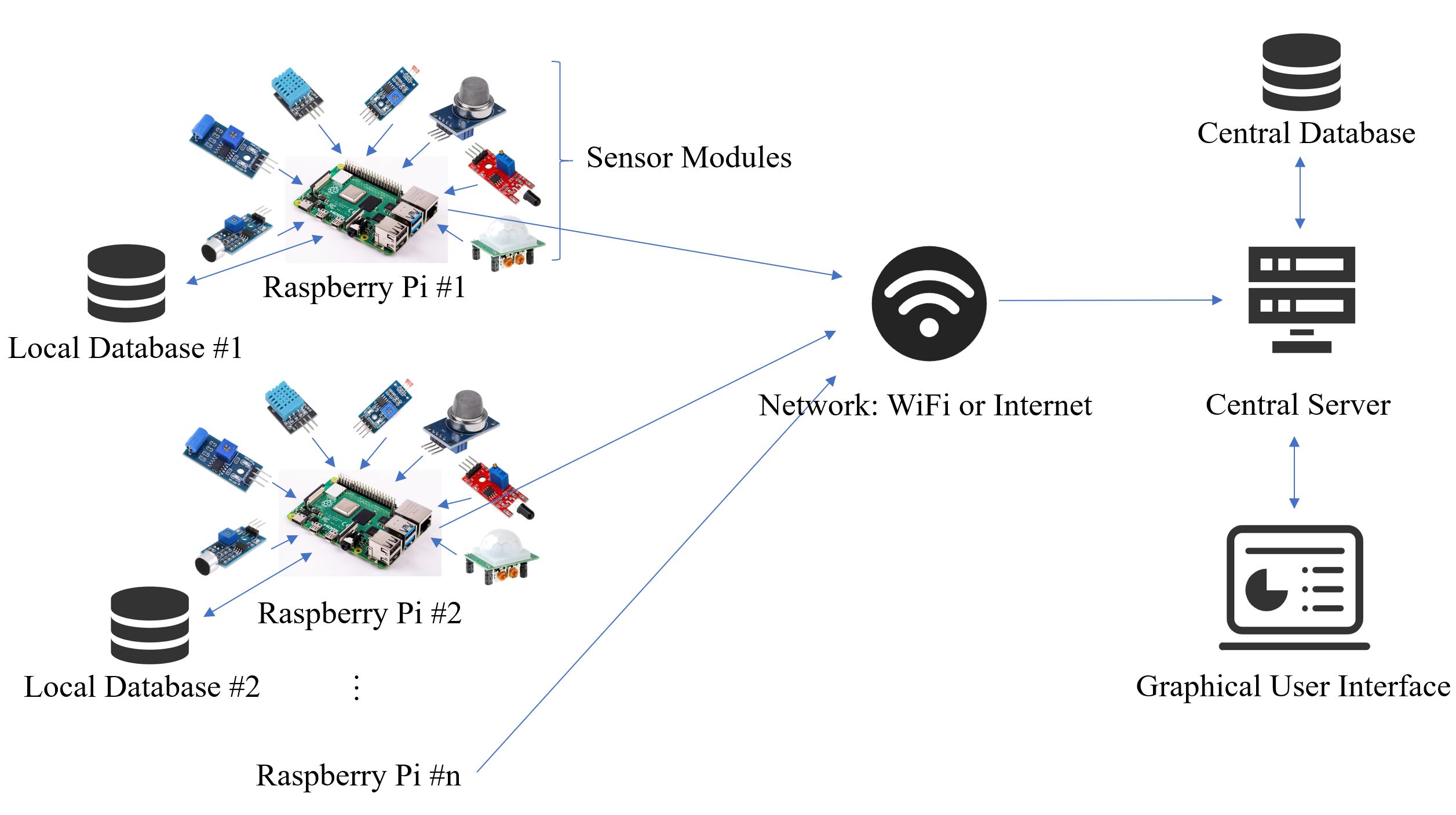}
\caption{The BDL system architecture.}
\label{fig:system}
\end{figure}

The BDL system, as shown in Figure \ref{fig:system}, consists of a central server and multiple individual sensing nodes, which are built around Raspberry Pis. The sensors in each sensing node capture real-time data from the environment and transfer the data to the connected Raspberry Pi, which has its own local database. After a certain interval, the Raspberry Pi communicates with the central server and upload the newly generated data to the central database. The GUI communicates with the central database to visualize data, and provides functions such as downloading (selected) data, downloading the error log, modifying the system by adding or removing sensing modules or sensing nodes. 

\begin{figure}[t]
\centering
\includegraphics[scale=0.25]{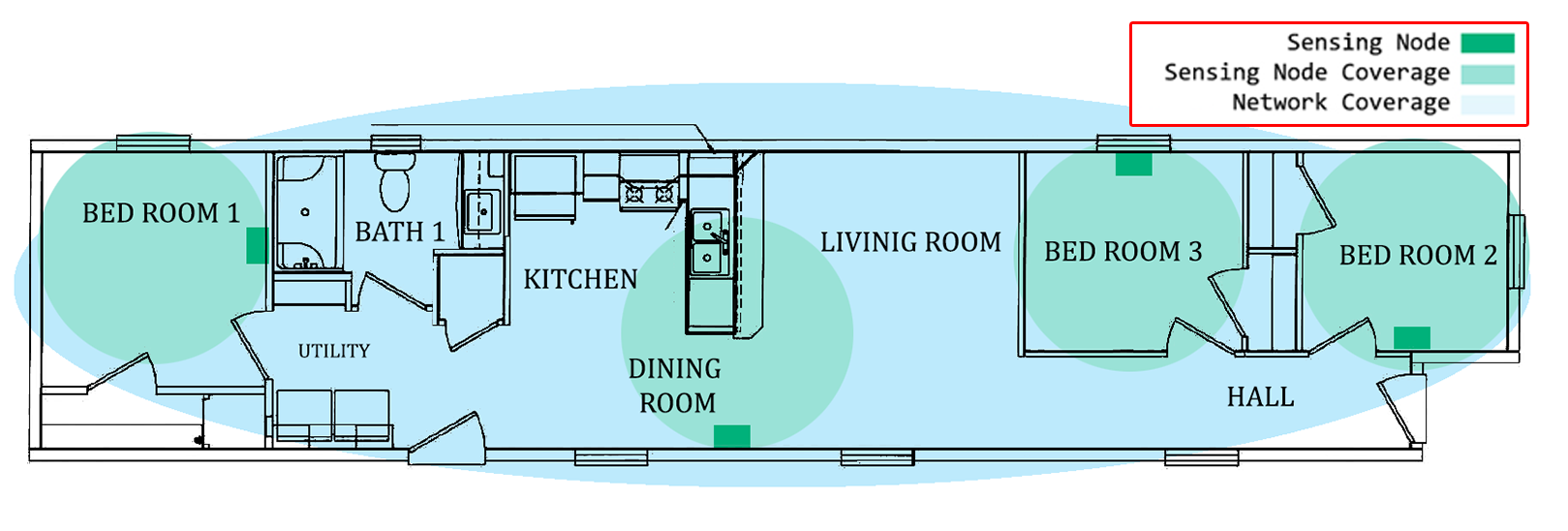}
\caption{An example of BDL deployment.}
\label{fig:implementation}
\end{figure}

Figure \ref{fig:implementation} illustrates an example of the BDL system deployment. The green portions represent the locations where individual sensing nodes collect environment data. The blue area denotes the area covered by the wireless network. The central server is also connected to this network either locally or through a live server. The sensing nodes are modular, and each node can have different sensors. They are portable and can be deployed anywhere with power and network connection. 

All software used in the system is open sourced, and both the database systems used, MariaDB and MySQL, are free. The system software including both server and sensing node has been made publicly available and can be found in the GitHub repository \cite{AnikGitHub2021}. The server is currently available for open access to help researchers and industry practitioners with the need of indoor environment sensing.




\subsection{Central Server}
\begin{figure}[t]
\centering
\includegraphics[width=\textwidth]{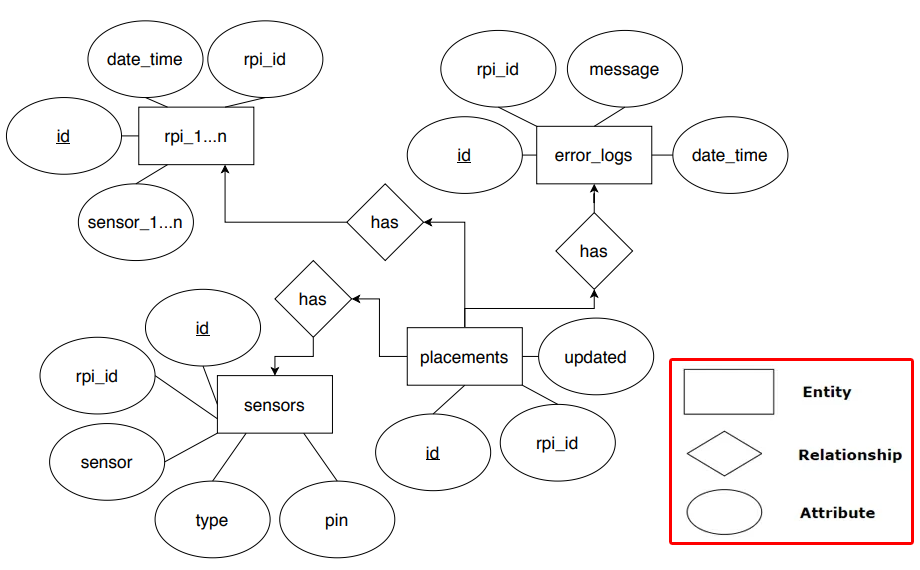}
\caption{Entity relationship diagram of the central database.}
\label{fig:erd_central}
\end{figure}

The central server of BDL is designed as a web-based system. The prototype uses PHP v7 as the server language. The front-end was written in HTML, CSS, and JavaScript. The system is designed to use Asynchronous JavaScript and XML (AJAX) to create a fast and dynamic GUI. JQuery and Chart.js libraries are used for data visualization. The GUI fetches data from the central database. The server is live and being hosted in a third party hosting sites with the domain: building-data-lite.com. Figure \ref{fig:erd_central} shows the entity relationship diagram of the central database, and it is a general overview of the database schema. Here the rectangular shapes denote the entities, the circular shapes denote the attributes of the entities and the rhombus shapes denote the relation between each entities. 

\begin{figure}[t]
\centering
\includegraphics[width=\textwidth]{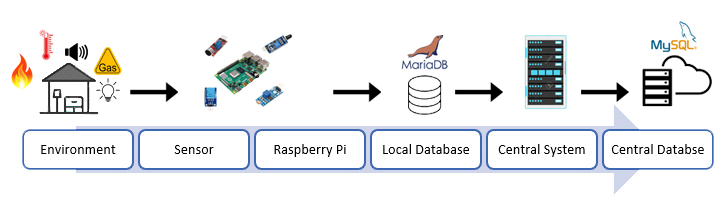}
\caption{Data flow in the BDL system.}
\label{fig:dataflow}
\end{figure}

Each sensing node has its local database system to store offline data. This is essential because there can be disruptions of communication between the sensing node and the central server. The sensing nodes use MariaDB for the local database. To keep data consistency, a temporary data file is being created every hour consisting of new records that have not been sent to the central server. The central server provides the information (for example, the timestamp of last synchronization) to make sure the data in this file are not redundant. In each synchronization, the file is transferred to the central server, and then, replaced by a new one to reduce space wastage. Approximately 60 records are present in each hourly generated file, but this number will be doubled for each occurrence of communication disruption. The central server performs a synchronization check of the received data before storing it into the central database. Figure \ref{fig:dataflow} illustrates the data flow in the BDL system. 

\subsection{Sensing Node}
\label{section:sensing_node}
\begin{figure}[t]
\centering
\includegraphics[width=\textwidth]{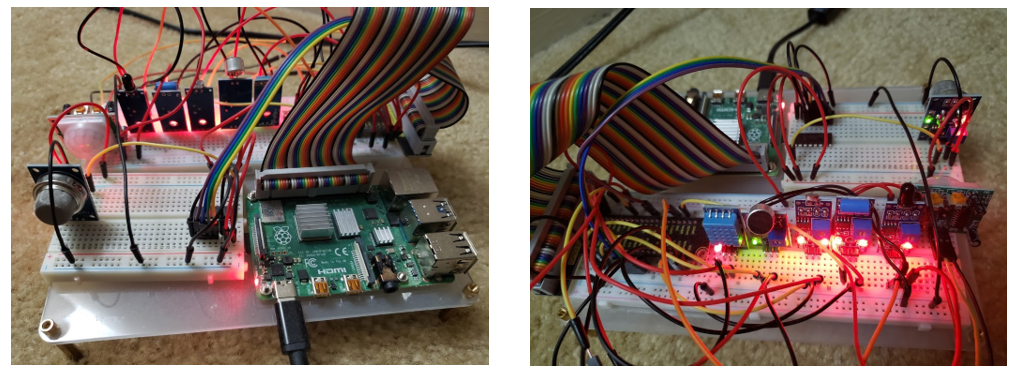}
\caption{Sensing node prototype version 1.}
\label{fig:prototype}
\end{figure}

In each sensing node, the Raspberry Pi connects with multiple sensor modules to collect data from the surroundings. Figure \ref{fig:prototype} shows the prototype of a sensing node. The sensors are connected via the GPIO ports of the Raspberry Pi, with a GPIO extension breadboard for easier deployment. The Raspberry Pi can take digital inputs only, which is a problem in incorporating analogue sensors, such as the light sensor and the gas sensors. To resolve this issue, the authors use an analogue-to-digital converter, MCP3008 \cite{mahmud2018complete}, to capture readings from the analogue sensors. The MCP3008 has 8 channels, which means it can connect up to 8 analogue sensors with the Raspberry Pi. The rest of the GPIO pins on-board will be able to connect one digital sensor each. The sensors have been categorized in three types depending on their implementation with the Raspberry Pi. Type 1 is the direct input sensors, for example, the light dependent resistance sensor (LDR) \cite{krishnamurthi2015arduino}. These sensors connect to the GPIO pins and Raspberry Pi can be directly read data from these pins. Type 2 refers to the sensors that provide a feedback signal to the connected pin when an event occurs, for example, the sound sensor \cite{saha2018raspberry, symon2017design} or the motion sensor \cite{sukmana2015prototype}. The third type of sensors do not share a common implementation pattern and needs to have a specific implementation code. The repository will be continuously updated with new sensors of this type for users to just select and use. Advanced users can add their custom code in the system for this type of sensors. The code repository \cite{AnikGitHub2021} includes a section to add custom codes for this sensor type. 

The range of each sensor is different and some sensors include a  potentiometer to tune the amplitude of readings or range of the sensor. As the system is designed to be of generic purpose, the range of a sensing node will depend on the sensors used in the particular node. For example, the default radius of a PIR motion sensor is 6 meters with 120 degree angle while a temperature sensor should capture the temperature of the air surrounding the sensor. The sensing node captures environment data through the connected sensor modules and stores them in its local database, which is designed based on the connected sensing modules. It captures one value from each sensor at every given time frame, for example, 60 seconds. The Raspberry Pi must have an operating system installed and include the required packages to run python scripts of the sensing program. The packages required to run the program is listed in a note with more instructions on the code repository \cite{AnikGitHub2021}. 

There are multiple options for the core of the sensing node design, such as ESP8266, ESP32, Nodemcu, Arduino, Pycom, and our selected board of Raspberry Pi. While the other boards are slightly cheaper than Raspberry Pi (model 4B), they are mostly microcontrollers and/with WiFi connectivity modules. These boards provide easier connectivity with low-level sensors but Raspberry Pi boards provide the functionalities of a complete computer system with operating system, memory, database, built-in WiFi module, HDMI, and USB extension ports. It is easier to program, collect and store data in a Raspberry Pi than the other boards, especially for a non-technical individual. 

\subsection{Database Configuration and Data Transmission}
\begin{figure}[t]
\centering
\includegraphics[scale=0.3]{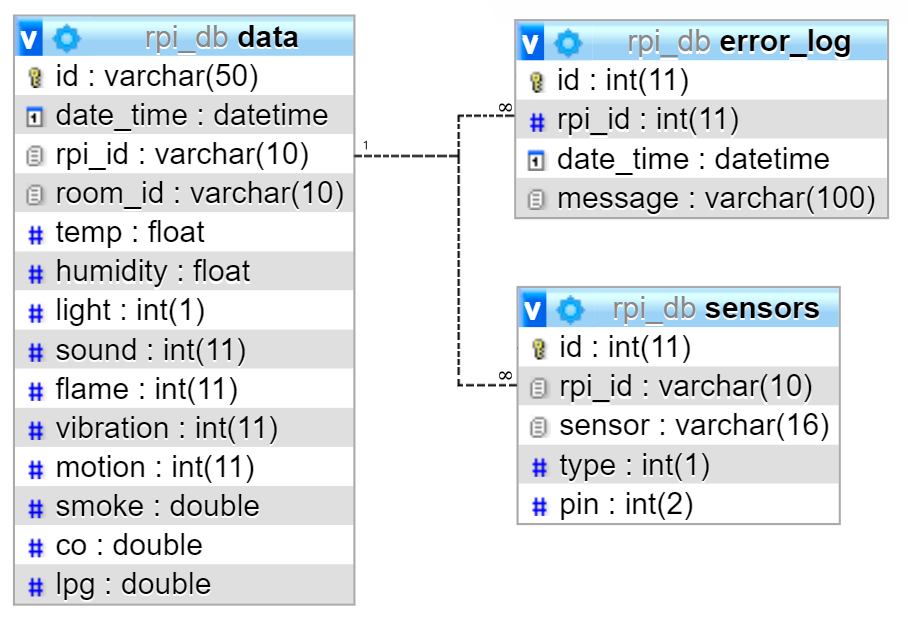}
\caption{Local database design.}
\label{fig:erd_local_db}
\end{figure}

There are two separate database designs in the BDL system. One is used in the sensing nodes (shown in Figure \ref{fig:erd_local_db}) and another in the central server (shown in Figure \ref{fig:erd_central}). The database share some common entities such as ``error\_logs'' and ``sensors''. The ``data\_storage'' table of the local database represents a specific Raspberry Pi table in the central database with the name ``rpi\_n'' where n represents the raspberry pi identification number. 

Specific implementation of the system will vary according to what sensor modules are used by each customized sensor nodes. The ``placements'' table records Raspberry Pi nodes presents in the system. It has a column ``updated'' which refers to changes made in the Raspberry Pi. It will be 1 if a sensor has been added in the system or removed from the system and 0 otherwise. The entity ``rpi\_1...n'' represents sensing nodes based on Raspberry Pi boards. A new table marked with an increasing identification number is created every time a new node is inserted in the system. Every node can have different set of sensors connected to it. The dynamic set of sensors makes changes of sensor columns ``sensor1\_1...n'' in the corresponding Raspberry Pi table. The sensor information of each sensing node is also recorded in the ``sensors'' table  which contain implementation specific information, such as sensor type and connecting pin number. Each record in the ``rpi'' table is denoted by a unique id generated from the combination of date\_time and rpi\_id. The date\_time and rpi\_id (Raspberry Pi ID) are string data and are present in separate columns to enable faster queries. The rest of the columns represent different information record related to the sensors, such as light presence, temperature, sound, etc. The central database contains additional tables to keep track of the sensing node placements and sensor modules. The ``error\_logs'' table contains exception message logs. The local database system used in each sensing module is MariaDB, which is usual for Raspberry Pi systems. The central database is created using MySQL. MariaDB and MySQL are similar in many aspects, and to keep concurrency between them does not take much effort. 

Data transmission in the system is done through post methods in Hypertext Transfer Protocol Secure (HTTPS) which is secure. Relaying on the security of HTTPS, external encryption and decryption mechanisms have not yet introduced in the current BDL system but there are plans to introduce those in future. The frequency of requests in the system will depend on the number of sensing nodes in the system and not the number of sensors in each node. This is because the a sensing node will send chunk of data collected from all its connected sensors in a preset interval which can be an hour or a day or a month. This practically eliminates the limit of server request bottleneck because it is very unlikely that multiple sensing nodes will hit the server at the exact same time. Even if such scenario occurs, the HTTPS server can handle at least 200 requests a minute. 

\subsection{Network Configuration}
The BDL system is designed to run in both local and global connectivity. For the global setup, the server needs to be running in a live domain. Many live domain hosting providers are capable of running PHP server. It is one of the most common server languages and easy to implement. After the server project folder has been uploaded to the server’s public HTML folder, the server can start running at the domain address. The domain address needs to be set in all sensing nodes, by the change of a single line of code. Each sensing connects to a wireless network. If a server is deployed in local mode (on one of the Raspberry Pis rather than on an independent computer), then all sensing nodes and the server needs to be connected to the same network. If working in the global mode, then each node and the server need to be connected to the Internet. 

The DBL system prototype 1 adopts the Raspberry Pi model 4B, and the prototype 2 and 3 adopts the Raspberry Pi Zero W because they all have built-in WiFi modules for wireless network connections. Data are transferred from the sensing nodes to the central server via the network, which can be WiFi or wireless Internet (4G, 5G, etc.). If the network connection is broken for some reason, the sensing nodes will keep working in offline mode, in which they will keep generating data and storing them in their local databases. The data will be synced whenever the connection is resumed. 

\subsection{Data Integration}
\begin{figure}[t]
\centering
\includegraphics[scale=0.6]{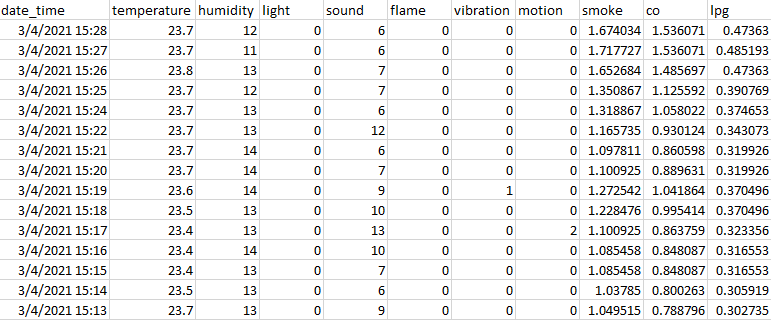}
\caption{Sample data format}
\label{fig:data_format}
\end{figure}

In each sensing node, the sensors output data to the connect Raspberry Pi either directly or through an analogue to digital converter (ADC), and the Raspberry Pi receives the data through the GPIO ports. The sensing node is programmed to run in an infinite loop to collect data continuously. In the prototypes, there is a 60 seconds interval between the readings, and it can be changed according to demand. The data record interval does not need to be manually adjusted and the user can view the data in different intervals on the GUI regardless of the input interval. The viewing interval however can be either equal or greater than the recorded interval, for example, if data is recorded every minute then the records cannot be viewed per 30 seconds but they can be viewed as per minute or as per 10 minutes. The Raspberry Pi collects all sensor readings each minute and merge them together to make a record. The record is then stored in the local database with a timestamp and other necessary information, such as location and the identifiers of sensor modules. If any of the sensor faults in providing data, a system exception is thrown. The sensing node records the exception details with the timestamp in a separate table in the local database. The data is stored in numeric format. Each sensor is different in regards to its readings, for example, the temperature sensor (DHT11) provides readings in degree centigrade while the humidity reading is in percentage. Figure \ref{fig:data_format} illustrates some data format sample from the sensors connected in the designed prototype. Here, each column from `temperature' to `lpg' refers to a sensor name and the records show the different type of values each sensor generates.  

In the prototype system, the sensing node is programmed to communicate with the central server once every hour (which can be changed). Each sensing node sends its identification string to the central server, and the server returns the ID of the latest stored data from the corresponding sensing node. After receiving that ID, the sensing node queries the local database for all the records inserted after that given ID. The local database then returns the newly inserted records to the sensing node. These are the records that are present in the sensing node but not yet inserted in the central database. The sensing node then creates a CSV file with these records and sends the generated file to the server. The central server receives the data and performs another synchronization check before inserting them to the central database.

\subsection{Graphical User Interface}

\begin{figure}[t]
\centering
\includegraphics[width=\textwidth]{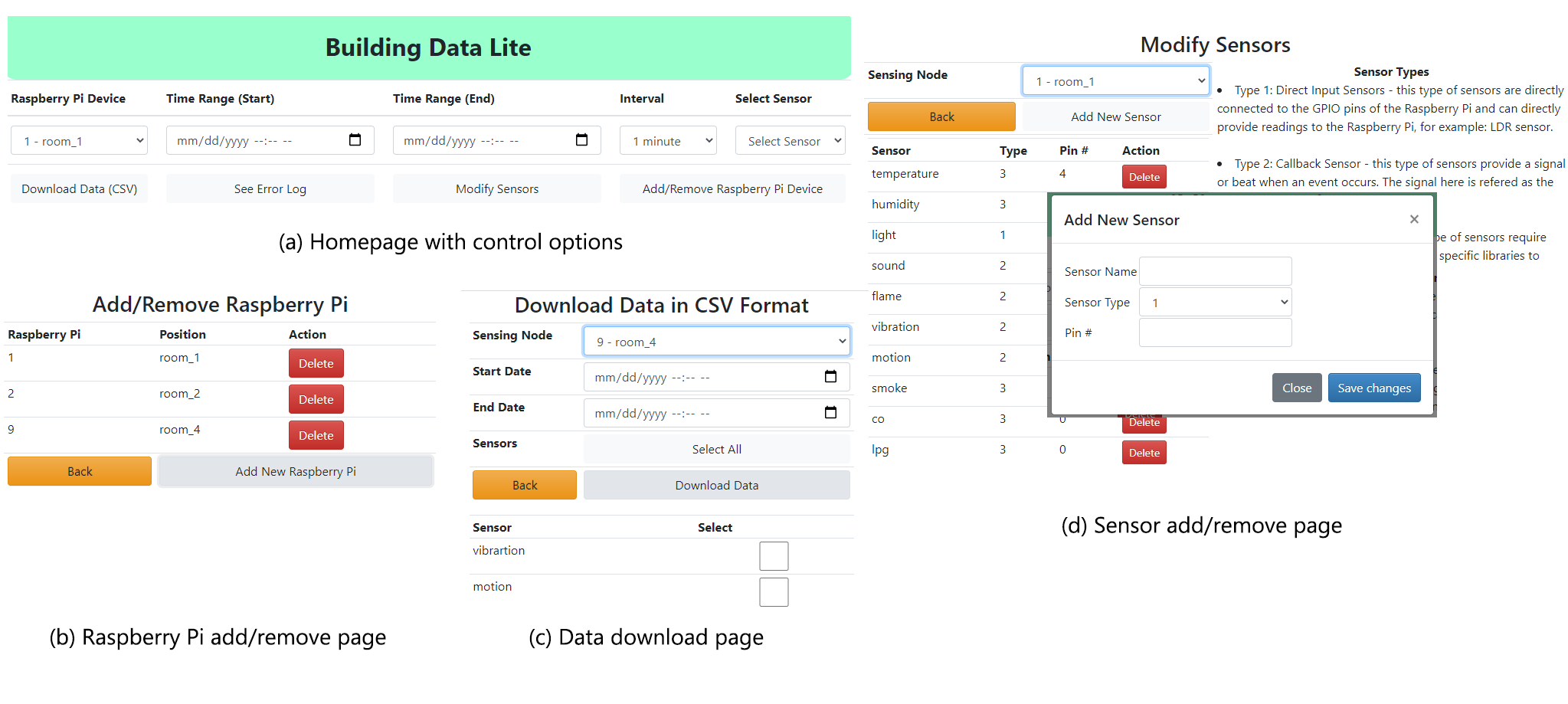}
\caption{BDL system's Graphical User Interface}
\label{fig:GUI}
\end{figure}

The BDL system's GUI has five functions, which are 1) data visualization, 2) data download, 3) error log download, 4) add or remove sensing nodes (Raspberry Pis), 5) modify sensor modules on each sensing node. Figure \ref{fig:GUI}(a) illustrates the home page, which updates dynamically on the value change of the control fields. To visualize the data, the user can specify sensing node(s), the time range, time interval, and sensor module(s). The data is generated by the sensors connected to the sensing nodes. The data is represented in a graph chart with the date-time in x-axis and sensor reading in the y-axis. The user can also download the numeric data in a CSV file through the data download page shown in Figure \ref{fig:GUI}(c). The user has the option to select the sensing node and sensors from the list of connected sensors. The system's error log collects error information from all the sensing nodes and it can be downloaded for each sensing node and specified time span, similar to data download page as shown in Figure \ref{fig:GUI}(c).

The BDL system architecture is designed in a way that it is capable of including or removing sensing nodes and sensors of each node as per the need. The user has to connect and disconnect the hardware manually, and the BDL system will adjust the code based on the user inputs from GUI. The GUI has the option to add/remove sensing nodes and sensors, which can be done by clicking the button. The user does not need to write any code to add or remove a sensing node or a sensor. After the operations on hardware, which are set up the Raspeberry Pi, and connecting or disconnecting sensor modules to it, the user can use the GUI to finish the system configuration, as shown in Figure \ref{fig:GUI}(b) and Figure \ref{fig:GUI}(d). 

The sensors are categorized in three types as described in section \ref{section:sensing_node}. The GUI offers a dropdown list for the user to select the sensor type when adding a new sensor. The user needs to specify the sensor name, type, and, connected pin in the GUI as shown in figure \ref{fig:GUI}(d). As mentioned in section \ref{section:sensing_node}, the sensors of type-3 require specific implementation code, to minimize the effort on the user, the GUI of BDL offers a list of supported sensors which the user can directly choose from. This list will be kept up to date to support latest sensors. The advanced users also have the option to write and add their custom code in the system for this type of sensors. The code repository\cite{AnikGitHub2021} includes a section to add custom codes for this sensor type.



\section{Prototype Deployment} \label{experiment}

In the BDL system, each sensing node is centered around its own Raspberry Pi, which is connected with a set of analogue and digital sensor modules. Any sensor that is capable of communicating through the General-Purpose Input Output (GPIO) ports of the Raspberry Pi system can be connected in this system. The first two prototypes of the BDL system were developed using the following sensors: DHT11 (temperature and humidity) sensor, light sensor, sound sensor, vibration sensor, motion sensor, MQ2 (smoke, natural gas and carbon mono-oxide) sensor. The third prototype version includes a sensor array named Enviro Plus \cite{enviro_plus}, which includes light, proximity, gas, and sound sensor.  Each sensing node is connected to the central server via either a WiFi network or a wireless Internet connection. The prototype 1 and 2 use the central server in local connectivity mode, in which the server and the sensing nodes need to be connected in the same wireless network to communicate. The prototype 3 uses the global mode, in which the server is implemented in a live server in the cloud, eliminating the need for being in the same wireless network with the sensing nodes. In this mode, each sensing node needs to be connected to the Internet via either wired or wireless connection. The Raspberry Pi model 4B used in the first prototype features both types of connections while the other two prototypes use Raspberry Pi Zero which uses the wireless communication.

\begin{figure}[t]
\centering
\includegraphics[width=\textwidth]{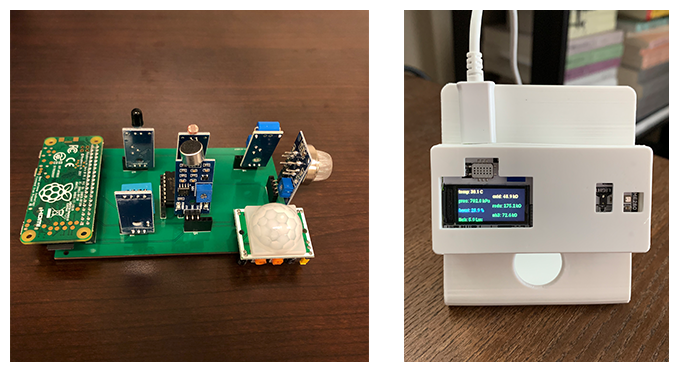}
\caption{Sensing node prototype version 2 (left) and version 3 (right).}
\label{fig:prototype_v2_v3}
\end{figure}

The sensing nodes of the BDL system are customizable and can be implemented in different ways. To demonstrate the different implementations, three separate variations of the sensing node were created and tested. While the server side remained the same, the physical hardware setup of the sensing nodes varied. Figure \ref{fig:prototype} shows the bread-board setup which is completely scalable with plug-in sensors (prototype version 1). Figure \ref{fig:prototype_v2_v3} (left) shows the Printed Circuit Board (PCB) version of the bread-board variant using a Raspberry Pi Zero, which is prototype version 2, and the third version of the prototype, which uses a Raspberry Pi Zero connected to an Enviro Plus sensor module, is shown in Figure \ref{fig:prototype_v2_v3} (right). 

\begin{figure}[h!]
\centering
\includegraphics[width=\textwidth]{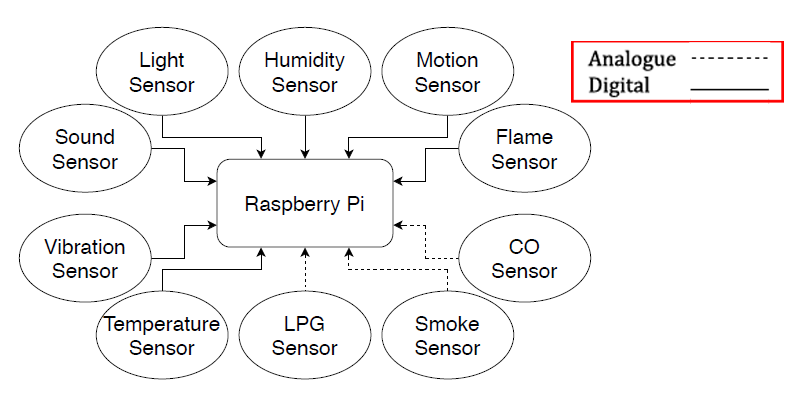}
\caption{Structure of the prototype's sensing node.}
\label{fig:node}
\end{figure}

Figure \ref{fig:node} shows the structure of one of the prototype sensing node (v1 \& v2). It contains seven sensors and provide 10 type of readings. On the figure, the solid lines represent the digital data and the dashed lines represent the analogue data. Prototype v3 has similar design with different set of sensors.

The sensors used in the prototype v1 \& v2 are listed as follows:

\begin{figure}[t]
\centering
\includegraphics[scale=0.8]{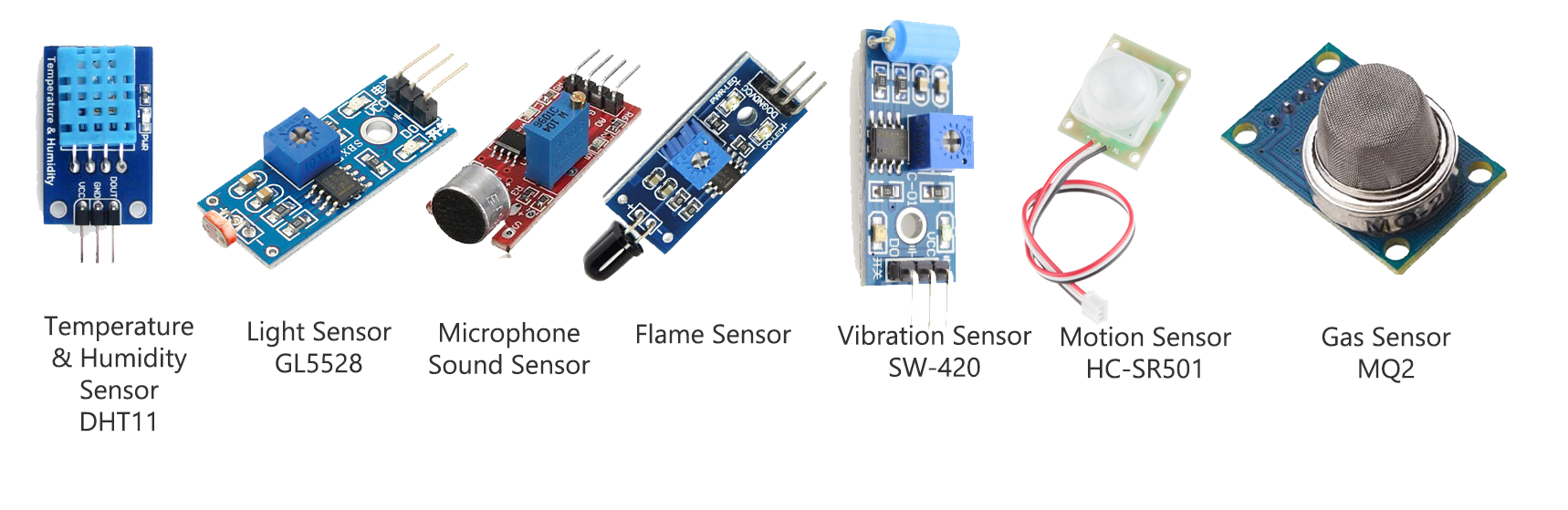}
\caption{Sensors used in sensing node prototype.}
\label{fig:sensors}
\end{figure}

\begin{itemize}
    \item Temperature and Humidity sensor (DHT11) : This sensor provides readings for temperature and humidity\cite{robotics2010dht11, uk2010humidity, AOSONG2010}. The system uses degree centigrade as the unit of temperature and percentage for the unit of humidity. It uses a capacitive humidity sensor and a thermistor to measure the surrounding air, and can generate new data every 2 seconds. 
    
    \item Light sensor (GL5528): The light sensor is one of the digital sensors. It uses photo-detector GL5528 to detect light intensity of the surrounding environment. The resistance of the sensor varies depending on the amount of light it is exposed to which changes the output voltage. It provides 0 when it detects light and 1 if it does not detect any light. The sensor has a tuning potentiometer (POT) for adjusting the reading threshold \cite{krishnamurthi2015arduino}. 
    
    \item Microphone Sound sensor: The sound sensor is also a digital sensor. When module in the intensity of the sound environment does not reach a threshold (set through the included variable resistance), the sensor will output a high signal. Otherwise, it will output a low signal. The sensor detects a change in sound amplitude and provides a reading \cite{saha2018raspberry, symon2017design}. This reading can be called a beat. The designed sensing node counts the number of beats in a time interval for the data record. 
    
    \item Flame sensor: This sensor detects the presence of flame nearby. It is sensitive to flame wavelengths between 760nm to 1100nm in infrared. It is a binary sensor and provides 1 for the presence of flame and 0 for otherwise. There is a tuning potentiometer to adjust the reading threshold of the sensor \cite{taha2018implementation}. 
    
    \item Vibration sensor (SW420): The vibration sensor can detect nearby vibration beyond a threshold defined by a potentiometer and provide a binary reading \cite{biansoongnern2016development}. The beat counting mechanism has been used to present the readings of this sensor (how many beats per minute). 
    
    \item Passive Infrared Motion sensor (HC-SR501): Passive Infrared motion sensor, also known as PIR sensor, records the intensity of motion detected nearby \cite{sukmana2015prototype}. It comes with two tuning potentiometers to adjust sensitivity and reading delay. The output goes high when objects enter the sensing range, and automatically returns to low when object leaves.
    
    \item Gas Sensor (MQ2): This is an analogue gas sensor that requires an additional analogue to digital converter circuit to incorporate this sensor with the Raspberry Pi. It can provide continuous readings of smoke, carbon monoxide (CO), and liquefied petroleum gas (LPG) \cite{heyasa2017initial}.
    
\end{itemize}

Prototype v3 uses a sensor array, Enviro Plus \cite{enviro_plus}, which packs light, proximity, temperature, humidity, gas, and sound sensors. The prototype's source code is shared through an open-source project published on by the authors \cite{AnikGitHub2021}. The wiring settings of each sensor is written in comments of the source codes. A GPIO extension for the Raspberry Pi has been used in the prototype development for the ease of implementation in a breadboard setup (v1). The digital sensors are directly connected to the GPIO pins of the Raspberry Pi. The analogue sensors are connected to the MCP3008 analogue-to-digital converter, which forwards a digital output to the Raspberry Pi for the respective analogue sensor connected. Figure \ref{fig:sensors} shows the different sensors used in the sensing node prototype. From the left, DHT11, light sensor, sound sensor, flame sensor, vibration sensor, motion sensor, and, MQ2 gas sensor. The version 2 uses the same setup in a printed circuit board setup, which makes the setup more compact. 

To demonstrate the flexibility and stability of the sensing nodes, a total of 30 nodes including all three variants have been developed and deployed which are continuously collecting data for over 4 months.





\section{A Case Study} \label{study}
To evaluate and demonstrate the capabilities of the BDL system, the research team has been conducting a case study. A total of 48 sensing nodes are being deployed and collecting data continuously in 12 households (four sensing nodes each, distributed in different rooms) of a affordable community in Richmond, VA. The data collected by the sensing nodes of BDL prototype 3 system are being transferred to a live cloud server.

\begin{figure}[h!]
\centering
\includegraphics[scale=0.15]{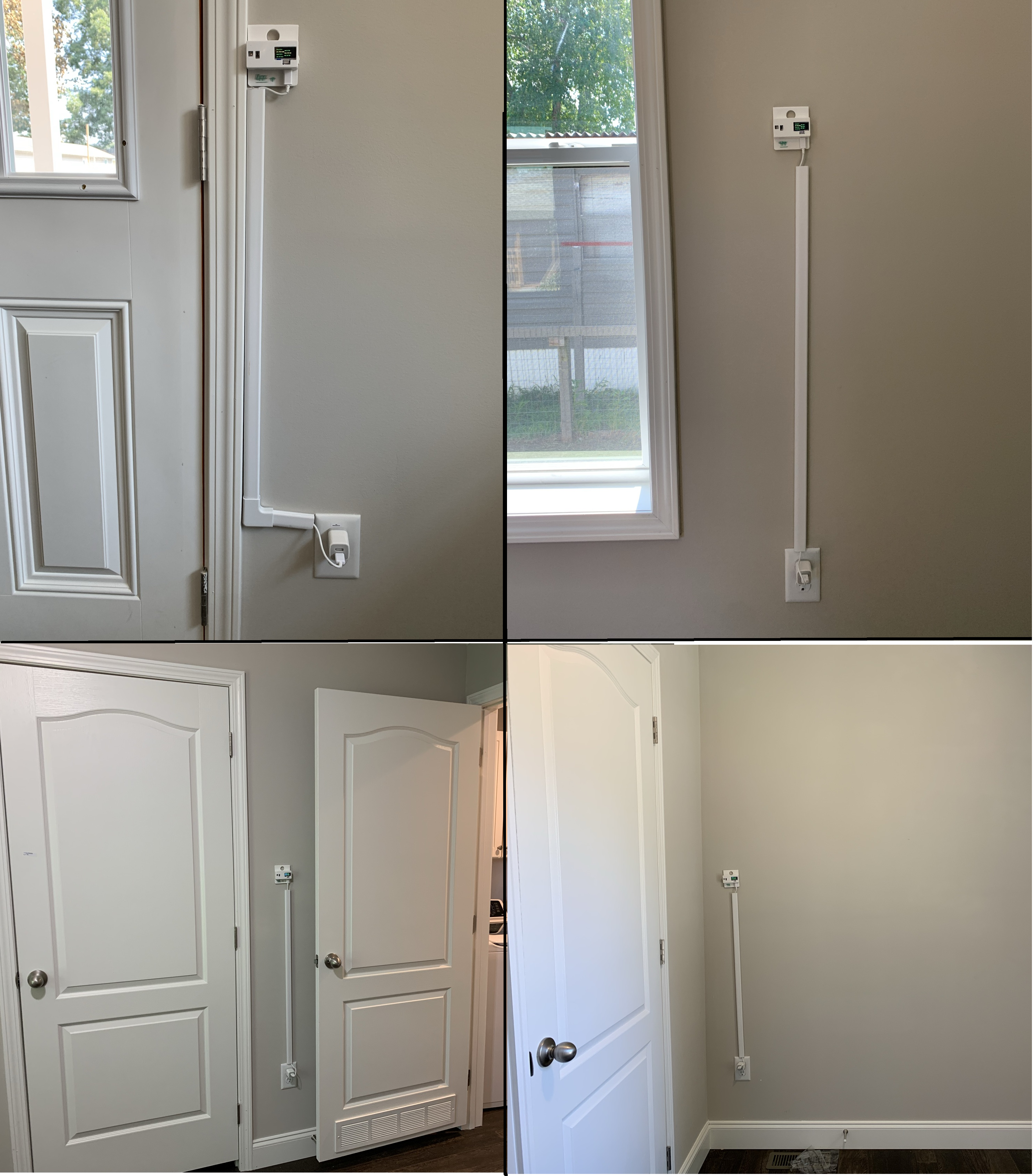}
\caption{Image a deployed sensing node.}
\label{fig:deployed_sensing_node}
\end{figure}

\begin{figure}[h!]
\centering
\includegraphics[scale=0.30]{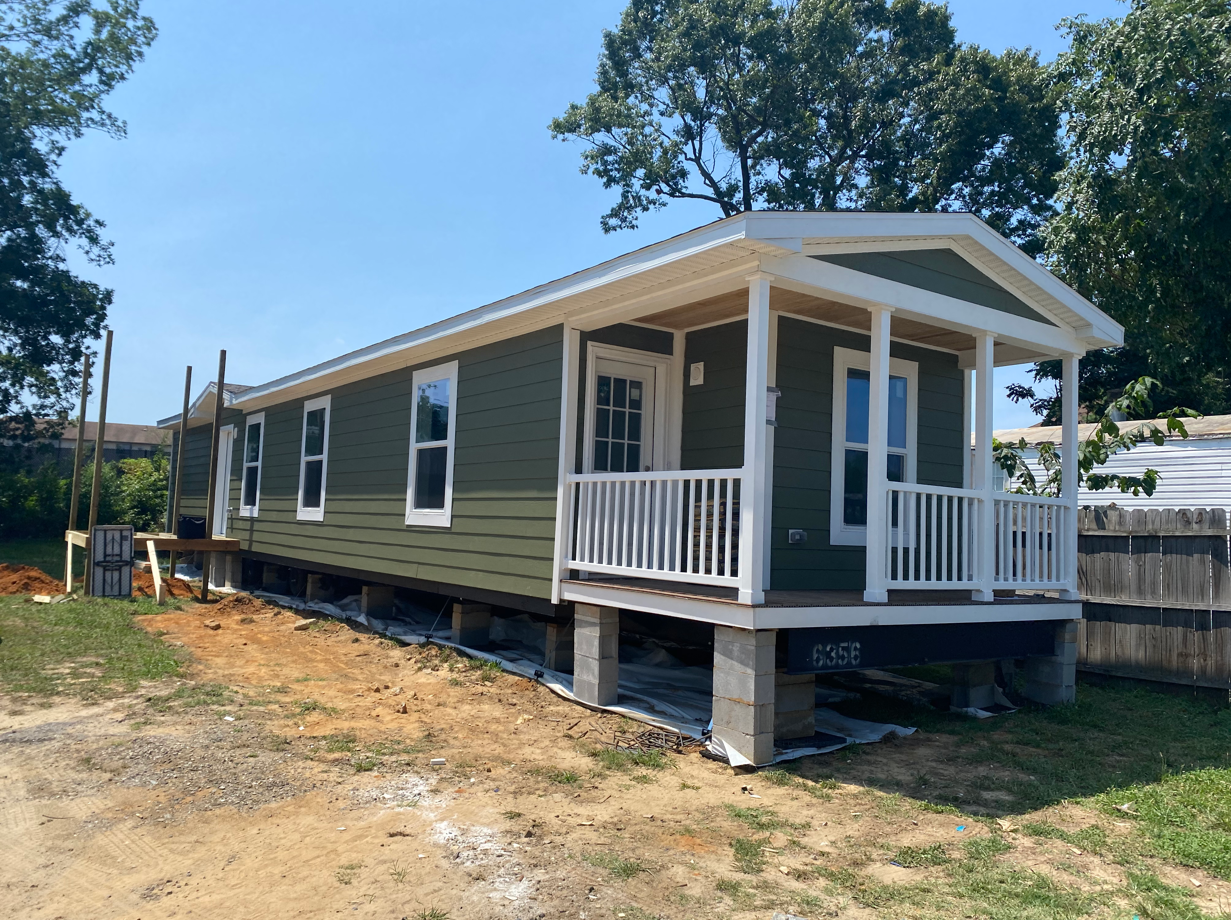}
\caption{A photo of a house in the case study.}
\label{fig:deployment_house}
\end{figure}

The typical sensing node deployment is illustrated in Figure \ref{fig:implementation}, in which one of the houses' floor plan and the locations of the devices are shown. Figure \ref{fig:deployed_sensing_node} shows a set of images about the deployed devices, and Figure \ref{fig:deployment_house} shows an image of a house in the case study. 

The research team use BDL prototype version 3 for the case study because it is more compact and, with the 3D-printed cases, it is more acceptable for the house occupants participating in this study. BDL v3 uses Raspberry Pi Zero with built-in WiFi module, connecting with the Enviro or Enviro Plus sensor bundles. The Enviro Plus bundle includes proximity, humidity, pressure, light, sound, and gas sensors. It is the advanced version of Enviro bundle, which includes all these sensors except for the gas sensor. The research team has to use some of the Enviro bundle because the global chip shortage is impacting the availability of the Enviro Plus bundle. The server side code is slightly modified so that both types of the sensor bundles can be installed in the system with just a single click. 

The deployed system's central server operates in global mode on an Apache server \cite{mockus2000case} of the Xampp module \cite{dvorski2007installing}. The server is established in a laptop with Microsoft Windows 10 as the operational system. The central database is built with MySQL \cite{shah2020hands}. Local WiFi networks are used to connect each house's sensing nodes to the cloud-based central server. The sensing nodes' local databases are built with MariaDB \cite{kenler2015mariadb}. If the server becomes unavailable in situations such as lost Internet connection, the data transfer will be cancelled. Instead of making a second communication attempt immediately, the sensing node will keep collecting more data and then repeats the data transfer process after another hour. In this process, the data may pile up in the sensing node, and the generated file size will keep increasing every hour but 60 records per hour will not make the file size larger than the transfer capability of the current network systems.



\begin{figure}[h!]
\centering
\includegraphics[scale=0.35]{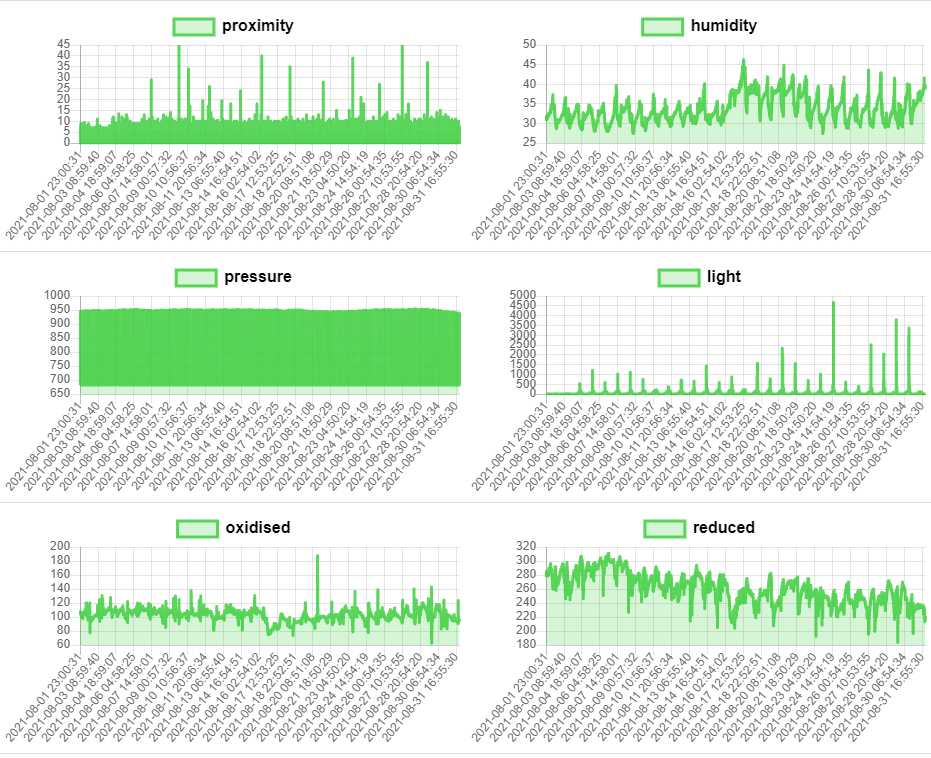}
\caption{Visualization of collected data from all sensors of a deployed sensing node.}
\label{fig:data_deployed}
\end{figure}
\begin{figure}[h!]
\centering
\includegraphics[scale=0.35]{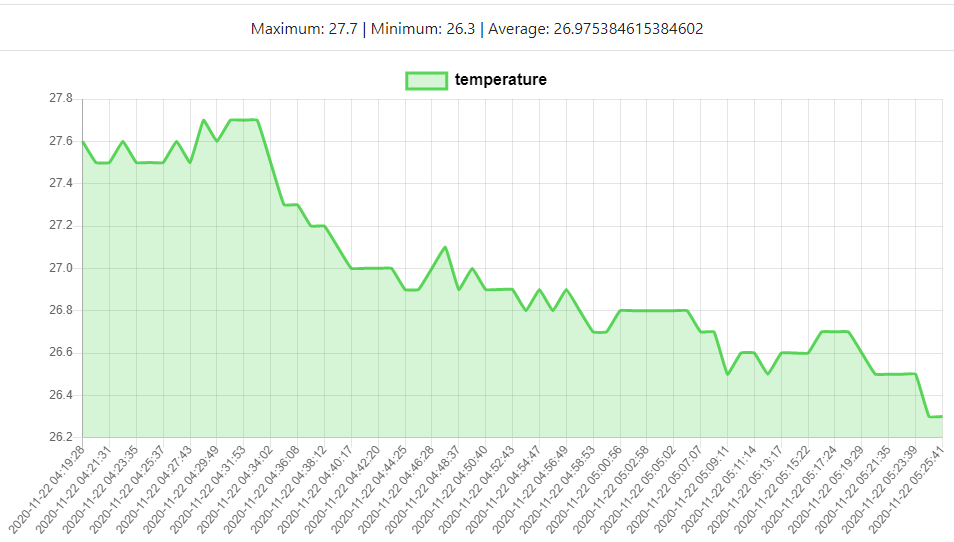}
\caption{Visualization of collected data of a single sensor.}
\label{fig:data_viz}
\end{figure}

The web-based GUI displays graphs of the data collected from the 12 houses and stored in the central server. With the proper selection of the control options (shown in Figure \ref{fig:GUI}a), the graph representation of the collected data can be presented on the website (building-data-lite.com). The system can show data graphs for all sensors of a sensing node on a single page, as Figure \ref{fig:data_deployed} shows, or the data graph of a particular sensor with additional detailed information such as minimum, maximum and average, as Figure \ref{fig:data_viz} shows. 

Figure \ref{fig:data_deployed} illustrates a truncated view of 6 sensor readings of a sensing node. The graph presents the collected data of this sensing node from 1st August 2021 to 31 August 2021. On the top row, the data of the proximity sensor and the humidity sensor are shown, and followed by the data of the pressure sensor and the light sensor on the second row, and then, the data of oxidised gases and reduced gases on the third row. The Figure \ref{fig:data_viz} shows the temperature data captured by the sensor node Raspberry Pi \#10 (rpi\_10). The minimum, maximum, and the average values are 27.7, 26.3 and, 26.98 centigrade degree, respectively. In all graphs, x-axis represents the date-time of data and the y-axis represents the reading of a particular sensor. 


The Enviro and Enviro Plus sensor board also include a LCD display. The BDL v3 sensing nodes utilize this display to provide the house occupants with a reading of the most recent data. 

The deployed devices in this case study are planned to collect data for at least one year. The data will be collected with full respect to the privacy of the tenants. These indoor environment data will be used for research studies related to low-income households, the performance of manufactured houses, and other related topics. 





\section{Discussion} \label{discussion}
\subsection{Challenges and Solutions}
This section discusses the technical challenges encountered by the authors and provides corresponding solutions.

\emph{Analog signal}. The Raspberry Pi, as the center of each sensing node, can only receive digital signals, but some of the sensor provide only analog signals (e.g. the gas sensor MQ2). This issue is solved by using an additional analogue-to-digital converter circuit. 

\emph{Binary sensors}. Some of the sensors provide only yes-no data. For example, the sound sensor gives a reading when detect a sound, but it cannot provide any measurement of the amplitude of the noise. The authors implemented a beat counting system to count the frequency of sounds during a time interval to achieve an understandable reading from the sensor.

\emph{Sudden crashes}. There were some sudden run-time crashes during the data collection phase. The problems were caused by faults in sensor readings, bad connection, overloaded access, and unavailability of server. Python's run-time exception handling feature was used to tackle these issues. Two types of exceptions were caught during this phase. The first one occurred due to a problem in sensor reading or sensor connection and the second one occurred due to server unavailability, longer wait time, or bad network connectivity. These exceptions were caught and logged in local database and later the log is transferred to the central database. 

\emph{Data synchronization}. Synchronization is important for the data continuity and integrity between each sensing module and the central server. The authors solved this issue using a two-fold solution combined together. First, a routine update mechanism between sensing node and central server was designed. Instead of transferring data records, we transferred files consisting of multiple records. Second, a record checking system was implemented in the sensing node to keep data consistency. The databases in both sensing nodes and the central server were restructured to reduce the number of entries (the database designs for the sensing node and the central server are shown by Figure \ref{fig:erd_central} and Figure \ref{fig:erd_local_db}, respectively).

\emph{File size limit}. The sensing nodes transmit the newly added records to the central database through a data file. If the transmission is not complete for some reason, such as the server being unavailable, the sensing node will wait another hour for the next transmission attempt. The file size increases as the number of unsent records piles up. The problem occurs when the file size exceeds 2 Megabytes, which is the default limit of file transfer in the local PHP server. The authors addressed this issue by manually overwriting the maximum transferable file size limit in the “php.ini” file. 

\subsection{Potential Use Cases}
The proposed BDL system is an innovative method to create the data infrastructure for existing buildings. It is also one step towards many potential smart building applications. Several example use cases are listed as follows:

\emph{Space utilization analysis}. The BDL system can be used to collect historical data of building space usage. These data can be used for space utilization analysis to deduce whether the space is being used properly or not; is it underused or is it overused; what is the indoor environment when the space is used, etc.

\emph{Building control system}. The BDL system can provide data to, or be merged with a building control system to enable automated building control. With sufficient data and analysis (such as machine learning on occupant behavior), the system can automatically control the indoor environment, such as light, temperature, humidity, ventilation, etc. 

\emph{Emergency management system}. The BDL system can be modified to become an emergency warning system. It can warn the building owner if it detects any abnormal characteristics in the indoor environment. The user can set a threshold for a particular reading to be marked as abnormal, or machine learning models can be developed to detect anomalies. 

\emph{Activity monitoring}. The BDL system can be used as a monitoring system for detecting unusual activities. It can provide data to, or be merged with a building security system that can warn the building owner that there is an unusual activity, such as burglary, in a particular building space.

\subsection{Limitations and Future Research}
Although the proposed BDL system's first prototype has been finished and experimented with, the system is still under development, and there are some limitations, which will be resolved or mitigated in the future. 

Some of the sensors require fine-tuning and calibration to provide a more accurate reading. The prototype has been calibrated, but for more accurate results, the system needs to be calibrated in different environmental conditions. The authors have plans to create more sensing nodes and deploy them in different locations. It will help us in tuning and calibrating the sensors for improved accuracy. 

The BDL system is capable of running in both local connection mode and global connection mode via local WiFi or Internet connection, respectively. The current prototype has only been tested in the local mode because of the lack of a live server. In the future, the system will be designed in a way that it will be able to switch between local mode and global mode.

\section{Conclusion} \label{conclusion}
This paper presents the research study that aims to establish a distributed wireless sensing network for collecting indoor environment data in existing buildings. This research contributes to the body of knowledge by proposing an innovative way for establishing a cost-effective, scalable, and portable IoT data infrastructure for indoor environment sensing. The authors first proposed the BDL system's overall design, and then developed three prototypes and conducted a case study with one of them to collect indoor environment data of 12 households of an affordable community. The data can be generated by the prototypes involve temperature, humidity, light, sound, flame, vibration, motion, smoke, carbon monoxide (CO), and liquefied petroleum gas (LPG).

The system's hardware is built around the mini computer Raspberry Pi and compatible sensor modules, and its software running on each Raspberry Pi is written in Python. The system includes a graphical user interface that enables the user to access, visualize, and download the data, and modify the sensing nodes and sensor modules as needed. The experiment has shown that the proposed system is functional, portable, and scalable. The system is also affordable with each sensing node costs only \$40 to \$60, and the software programs used are free. The authors have published the system prototype's source code on GitHub \cite{AnikGitHub2021}. 

This paper also discusses the challenges during the system development, corresponding solutions, potential use cases, and limitations of the current BDL system. The system has the capabilities to provide indoor environment data for research and applications in areas such as smart building, facility management, digital twin, and internet of things (in the built environment). Although the proposed BDL system's first three prototypes have been developed and experimented with, the system is still under development and require more adjustments and modifications to perform better. The authors plan to keep working on the system and conduct more studies with it in the future.

\section*{Funding}
This research was funded by NSF-1845446 and NSF-1929701 grants. 

\section*{Acknowledgements}
 We thank Zack Miller, Marion Cake, and Madeline Petrie at project:HOMES for supporting the case study. We thank National Science Foundation, Virginia Housing Development Authority, and Virginia Center for Housing Research for the research funding  support.

\bibliography{elsarticle-template}

\end{document}